\documentclass[prl,twocolumn,showpacs,preprintnumbers,amsmath,amssymb,superscriptaddress]{revtex4}

\usepackage{graphicx}
\usepackage{dcolumn}
\usepackage{bm}
\usepackage{tabularx}

\begin{document}

\preprint{}

\title{Spectroscopy of proton-rich $^{66}$Se up to $J^{\pi}$~=~6$^+$: isospin-breaking effect in the \textit{A}~=~66 isobaric triplet}

\author{P.~Ruotsalainen}
\email{panu.ruotsalainen@jyu.fi}
\affiliation{Department of Physics, University of Jyv\"askyl\"a, P. O. Box 35, FI-40014, Jyv\"askyl\"a, Finland}

\author{D.G.~Jenkins}
\affiliation{Department of Physics, University of York, Heslington, York YO10 5DD, UK}

\author{M.A.~Bentley}
\affiliation{Department of Physics, University of York, Heslington, York YO10 5DD, UK}

\author{K.~Auranen}
\affiliation{Department of Physics, University of Jyv\"askyl\"a, P. O. Box 35, FI-40014, Jyv\"askyl\"a, Finland}

\author{P.J.~Davies}
\affiliation{Department of Physics, University of York, Heslington, York YO10 5DD, UK}

\author{T.~Grahn}
\affiliation{Department of Physics, University of Jyv\"askyl\"a, P. O. Box 35, FI-40014, Jyv\"askyl\"a, Finland}

\author{P.T.~Greenlees}
\affiliation{Department of Physics, University of Jyv\"askyl\"a, P. O. Box 35, FI-40014, Jyv\"askyl\"a, Finland}

\author{J.~Henderson}
\affiliation{Department of Physics, University of York, Heslington, York YO10 5DD, UK}

\author{A.~Herz\'{a}\v{n}}
\affiliation{Department of Physics, University of Jyv\"askyl\"a, P. O. Box 35, FI-40014, Jyv\"askyl\"a, Finland}

\author{U.~Jakobsson}
\affiliation{Department of Physics, University of Jyv\"askyl\"a, P. O. Box 35, FI-40014, Jyv\"askyl\"a, Finland}

\author{P.~Joshi}
\affiliation{Department of Physics, University of York, Heslington, York YO10 5DD, UK}

\author{R.~Julin}
\affiliation{Department of Physics, University of Jyv\"askyl\"a, P. O. Box 35, FI-40014, Jyv\"askyl\"a, Finland}

\author{S.~Juutinen}
\affiliation{Department of Physics, University of Jyv\"askyl\"a, P. O. Box 35, FI-40014, Jyv\"askyl\"a, Finland}

\author{J.~Konki}
\affiliation{Department of Physics, University of Jyv\"askyl\"a, P. O. Box 35, FI-40014, Jyv\"askyl\"a, Finland}

\author{M.~Leino}
\affiliation{Department of Physics, University of Jyv\"askyl\"a, P. O. Box 35, FI-40014, Jyv\"askyl\"a, Finland}

\author{G.~Lotay}
\affiliation{Department of Physics and Astronomy, University of Edinburgh, Edinburgh, EH9 3JZ, UK}

\author{A.J.~Nichols}
\affiliation{Department of Physics, University of York, Heslington, York YO10 5DD, UK}

\author{A.~Obertelli}
\affiliation{CEA Saclay, IRFU/Service de Physique Nucl\'{e}aire, F-91191 Gif-sur-Yvette, France}

\author{J.~Pakarinen}
\affiliation{Department of Physics, University of Jyv\"askyl\"a, P. O. Box 35, FI-40014, Jyv\"askyl\"a, Finland}

\author{J.~Partanen}
\affiliation{Department of Physics, University of Jyv\"askyl\"a, P. O. Box 35, FI-40014, Jyv\"askyl\"a, Finland}

\author{P.~Peura}
\affiliation{Department of Physics, University of Jyv\"askyl\"a, P. O. Box 35, FI-40014, Jyv\"askyl\"a, Finland}

\author{P.~Rahkila}
\affiliation{Department of Physics, University of Jyv\"askyl\"a, P. O. Box 35, FI-40014, Jyv\"askyl\"a, Finland}

\author{M.~Sandzelius}
\affiliation{Department of Physics, University of Jyv\"askyl\"a, P. O. Box 35, FI-40014, Jyv\"askyl\"a, Finland}

\author{J.~Sar\a'en}
\affiliation{Department of Physics, University of Jyv\"askyl\"a, P. O. Box 35, FI-40014, Jyv\"askyl\"a, Finland}

\author{C.~Scholey}
\affiliation{Department of Physics, University of Jyv\"askyl\"a, P. O. Box 35, FI-40014, Jyv\"askyl\"a, Finland}

\author{J.~Sorri}
\affiliation{Department of Physics, University of Jyv\"askyl\"a, P. O. Box 35, FI-40014, Jyv\"askyl\"a, Finland}

\author{S.~Stolze}
\affiliation{Department of Physics, University of Jyv\"askyl\"a, P. O. Box 35, FI-40014, Jyv\"askyl\"a, Finland}

\author{J.~Uusitalo}
\affiliation{Department of Physics, University of Jyv\"askyl\"a, P. O. Box 35, FI-40014, Jyv\"askyl\"a, Finland}

\author{R.~Wadsworth}
\affiliation{Department of Physics, University of York, Heslington, York YO10 5DD, UK}


\begin{abstract}
Candidates for three excited states in the $^{66}$Se have been identified using the recoil-$\beta$ tagging method together with a veto detector for charged-particle evaporation channels. These results allow a comparison of mirror and triplet energy differences between analogue states across the \textit{A}~=~66 triplet as a function of angular momentum. The extracted triplet energy differences follow the negative trend observed in the $f_{7/2}$ shell. Shell-model calculations indicate a continued need for an additional isospin non-conserving interaction in addition to the Coulomb isotensor part as a function of mass.
\end{abstract}
\pacs{21.10.Sf, 21.10.Re, 21.60.Cs, 27.50.+e}    
                       
\maketitle
The building blocks of the nucleus, \textit{i.e.} the protons and the neutrons are conventionally regarded as two different particle species differing in charge and slightly in mass. However, as these particles are affected similarly by the strong nuclear force, they can be viewed as two different quantum states of a generic particle, the nucleon. This approach leads to the concept of isospin in which nucleons are distinguished by a z-projection \textit{T}$_z$ of the isospin quantum number \textit{T}. The isospin representation simplifies the treatment of the two-body nucleon-nucleon interaction and the classification of nuclear states. Isospin symmetry implies that for mirror nuclei, which have the same mass, but where the number of protons and neutrons is interchanged, the resulting analogue states with the same \textit{T} are degenerate. However, this degeneracy is lifted by isospin non-conserving (INC) forces, which lead to the mirror energy differences (MED)~\cite{bentley} evaluated as:
\begin{equation} 
\text{MED}_{J,T} = E^{*}_{J,T,T_{z} = -1} - E^{*}_{J,T,T_{z} = +1}. \label{MED}
\end{equation}
The MED relate to isovector energy differences; if the nuclear interaction was charge-symmetric in the absence of the Coulomb force, then the MED ought to be zero. In practice, it is found that the MED vary as a function of angular momentum on an energy scale of around $\sim$100~keV. Even on the assumption of perfect symmetry of the wave functions for isobaric analogue states, calculating the MED for a specific case can be complex. In addition to the expected two-body Coulomb effects, contributions to the MED are found from monopole effects such as single-particle Coulomb shifts, the electromagnetic spin-orbit interaction, and changes in radius or shape as a function of spin. In cases of weak binding, the breakdown of symmetry can also lead to further effects such as Thomas-Ehrman shifts~\cite{ehrman,thomas}. Where mirror states are well bound, there has been considerable success in calculating the MED and a good correspondence is found with experiment for nuclei in the $f_{7/2}$ shell~\cite{bentley}.\par

Analogue states in pairs of mirror nuclei are subsets of complete isobaric multiplets - \textit{i.e.} sequences of isobars where states are characterised by the same \textit{T}. A simple case is that of \textit{T}~=~1 triplets, in nuclei with $T_z=(N-Z)/2=0,\pm1$  where, in addition to the MED, the triplet energy differences (TED)~\cite{bentley} may be evaluated:
\begin{equation} 
\text{TED}_{J,T} = E^{*}_{J,T,T_{z} = -1} + E^{*}_{J,T,T_{z} = +1} - 2E^{*}_{J,T,T_{z}=0}. \label{TED}
\end{equation}
The TED are isotensor energy differences and probe a different aspect of the two-body interaction. They are sensitive to charge-dependent effects since they reflect the difference between the average of the proton-proton (pp) and neutron-neutron (nn) interactions and the neutron-proton (np) interaction. The TED have a special property that make them particularly attractive to study. That is, the TED are not expected to be strongly influenced by the single-particle contributions described earlier, but are instead especially sensitive to the details of the isotensor (multipole) interactions. At a fundamental level, these interactions may have one or two possible origins: a Coulomb interaction and/or a nuclear INC interaction, thus the TED have the capability to shed light on the balance between these terms.\par

Extensive information on the MED and TED exists for the \textit{sd} shell, where the relevant nuclei lie close to or on the line of stability (for most recent example see Ref.~\cite{jenkins_A23}). Over the last fifteen years, information on low-lying excited states has been gathered in the $f_{7/2}$ shell, allowing the MED and TED to be studied for the \textit{A}~=~46~\cite{garrett_46Cr} and \textit{A}~=~54~\cite{gadea_54Ni} triplets. In the upper \textit{fp} shell, however, the experimental information is extremely limited for odd-odd \textit{N}~=~\textit{Z} nuclei between $^{56}$Ni and $^{100}$Sn and almost non-existent for \textit{T}$_z$~=~-1 nuclei. This is undoubtedly due to the low production cross-sections for such nuclei as they lie very far from the line of stability. Here, the nuclear structure is expected to become significantly more complex with more orbitals involved. In addition, there is evidence of a sudden structural change when going towards the mass \textit{A}~=~70~$-$~80 region and the phenomena of shape coexistence, driven by the increasing occupancy of the $g_{9/2}$ orbital, is observed~\cite[and references therein]{hasegawa}. Aside from the pure nuclear structure interest, a deeper understanding of Coulomb and other INC effects across medium-mass \textit{T}~=~1 triplets may impact on related areas of physics including standard model tests~\cite{hardy-prc} and nuclear astrophysics~\cite{schatz}. For these reasons, it would be of high interest to pursue the TED and MED investigations beyond $^{56}$Ni. Recently, Obertelli {\em et al.,}~\cite{obertelli} identified the 2$^{+}$ state in $^{66}$Se in a study of two-nucleon removal from a secondary beam of $^{68}$Se at MSU. This constitutes the only definite identification of an excited 2$^+$, \textit{T}~=~1 state in the upper \textit{fp} shell (although tentative transitions have been reported for $^{62}$Ge~\cite{rudolph_62Ge}). In this paper, we present the observation of the $2^+$, $4^+$ and $6^+$ states in $^{66}$Se, allowing the most complete TED study to date above the $f_{7/2}$ shell.\par

In recent years the study of exotic nuclei has been driven by advances in experimental sensitivity concomitant with advances in detection technology. An example of a technique which can extract the signal of the exotic nucleus with exquisite sensitivity is recoil-decay tagging (RDT)~\cite{RDT1,RDT2}. This technique exploits the characteristic decay properties of the nucleus of interest to identify it at the focal plane of a recoil separator and then tag the associated $\gamma$ rays. Recently, RDT has been developed from its initial focus on alpha-decaying nuclei to be more broadly applicable. A challenging extension has been to $\beta$-decaying nuclei since, in general, $\beta$ decay does not provide a unique tag due to the three-body nature of the decay. In some special cases, however, $\beta$ decay can be used, where the character of the decay is Fermi superallowed. Here, the short half-lives ($\sim$100~ms) and high end-point energies ($\sim$10~MeV), in comparison the other neighbouring nuclei with end-point energies of $\sim$3~MeV and half-lives from seconds to hours, provide defining characteristics which can be exploited by correlating positrons with recoils implanted at the focal plane of a recoil separator. This technique, entitled recoil-$\beta$ tagging (RBT), is suitable for studying exotic proton-rich nuclei and was first demonstrated for $^{74}$Rb~\cite{steer} at the University of Jyv\"askyl\"a (JYFL). This initial work has latterly been extended at JYFL to the previously unknown case of $^{78}$Y~\cite{singh} and recently, provided additional information on excited states in $^{66}$As~\cite{rudi}. To reach the most exotic nuclei on the proton-rich side of the \textit{N}~=~\textit{Z} line in fusion-evaporation reactions, the experimental sensitivity needs to be still increased. These nuclei are also associated with pure neutron emission amidst a dominant background of charged-particle evaporation channels. In the present work, a charged-particle veto detector has been developed to suppress the reaction channels associated with proton and alpha evaporation. The effectiveness of this methodology is demonstrated with the important case of $^{66}$Se, in which the first excited 2$^+$ state has recently been identified~\cite{obertelli}.\par

The experiment was performed at JYFL utilising the K-130 cyclotron, which provided a $^{28}$Si beam at an energy of 75~MeV. The beam bombarded a $^{nat}$Ca target, rolled to a thickness of 0.65~mg/cm$^2$, with an average intensity of 3~pnA for 36 hours. Gamma rays were detected at the target position by the JUROGAMII array consisting of 24 clover~\cite{Clover} and 10 tapered~\cite{Phase1,GASP} Compton-suppressed germanium detectors with a total efficiency of 5.5~\% at 1.33~MeV. A new veto device, UoYtube (University of York tube), consisting of 96 CsI(Tl) crystals read out by photodiodes, was installed at the target position~\cite{jack}. Fusion recoils were separated from the beam by the gas-filled recoil separator RITU~\cite{Leino,saren}. Further identification of the recoils was performed in the GREAT~\cite{GREAT} spectrometer, located at RITU's focal plane, where the recoils were finally implanted in a pair of adjacent 700~$\mu$m thick double-sided silicon strip detectors (DSSD). The GREAT spectrometer also included a big segmented clover-, two JUROGAMII clover- and planar germanium detectors, which were mounted around the DSSD to observe delayed $\gamma$ rays. In addition, the planar detector in combination with the DSSD served as a $\Delta$\textit{E}-\textit{E} telescope for $\beta$ particles. Data were collected with the triggerless total data readout (TDR)~\cite{TDR} acquisition system and analysed with the GRAIN~\cite{GRAIN} software.\par

\begin{figure}
\includegraphics[width=0.48\textwidth]{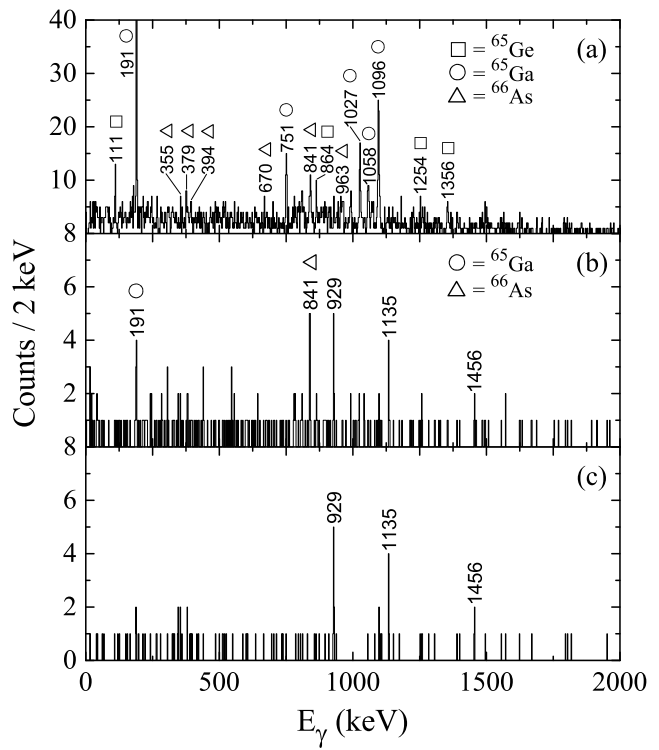}
\caption{\label{fig1}(a) Recoil-$\beta$ tagged $\gamma$-ray spectrum requiring detection of a 0.5~$-$~10-MeV $\beta$ particle in the planar detector in coincidence with the DSSD event that occurred in the same pixel as the recoil within a time window of 106~ms. (b) Same as (a) but with charged-particle suppression. (c) Same as (b) but with an additional veto condition obtained from delayed $\gamma$ rays detected at the focal plane of RITU (see text for details).}
\end{figure}

The identification of $^{66}$Se $\gamma$ rays is facilitated by its Fermi superallowed $\beta$-decay nature and by the fact that $^{66}$Se is produced via two-neutron evaporation, while the other products involve emission of at least one charged particle. With these features in mind, a step-wise procedure was followed to search for $\gamma$ rays originating from $^{66}$Se. In the first instance, the RBT method was applied by correlating 0.5~$-$~10-MeV $\beta$ particles to recoils within a correlation time of 106~ms [$\approx 3 \times \text{t}_{1/2}({}^{66}\text{Se})$~\cite{blank_66Se}]. Figure~\ref{fig1}(a) shows the observed $\gamma$ rays when these tagging conditions are applied. As expected, transitions from $^{66}$As are identified along with contaminants such as $^{65}$Ga and $^{65}$Ge corresponding to 3p and 2pn channels, respectively. Since $^{66}$Se is produced via 2n evaporation, the analysis may be refined by removing recoils from the correlations if an associated charged particle had been detected in UoYtube. Figure~\ref{fig1}(b) demonstrates how successful this approach is - the suppression of charged-particle evaporation leaves five peaks at 191~keV, 841~keV, 929~keV, 1135~keV and 1456~keV, where the first two can be associated with $^{65}$Ga and $^{66}$As, respectively. The $\gamma$ rays, detected in the focal plane in delayed coincidence with a recoil implantation or in prompt coincidence with $\beta$ decay, can be utilised as an additional veto condition. The $\beta$ decay of $^{65}$Ga feeds excited states in $^{65}$Zn, which are de-excited by various $\gamma$ rays such as 61-keV and 115-keV transitions. These $\gamma$ rays can be observed in the focal plane germanium detectors in prompt coincidence with the $\beta$ decay with high efficiency ($\epsilon_{61~\text{keV}}$~=~23~\%, $\epsilon_{115~\text{keV}}$~=~20~\%). If at least one of these $\gamma$ rays has been detected, the associated recoil event preceding the $\beta$ decay is omitted from the tagging process, hence it is not correlated with the prompt $\gamma$-ray transitions. The 841-keV transition feeds an isomeric state in $^{66}$As de-excited by a 114-keV transition, which is followed by the emission of eight other $\gamma$ rays~\cite{rudi}. In this case, if any of these nine transitions is observed in the focal plane germanium detectors, the associated $^{66}$As recoil is removed from the tagging process. When all known delayed $\gamma$ rays, either following recoil or $\beta$ decay, are considered, the resulting total focal plane veto efficiency is sufficiently high resulting in a tagged $\gamma$-ray spectrum with three lines as shown in Fig.~\ref{fig1}(c). The $\gamma$ ray at 929(2)~keV de-excites the 2$^+$ state in $^{66}$Se, since it is consistent with the transition energy of 929(7)~keV reported in Ref.~\cite{obertelli}. The other two peaks at 1135(2)~keV and 1456(2)~keV are assigned to de-excite the 4$^+$ and 6$^+$ levels, respectively, as the observed pattern represents a typical spectrum of the strongest yrast transitions in an even-even nucleus. The relative intensities of the transitions depopulating the 2$^+$ [I$_{929~\text{keV}}$=100(37)], 4$^+$ [I$_{1135~\text{keV}}$=70(43)] and 6$^+$ [I$_{1456~\text{keV}}$=50(34)] states are very similar to the corresponding yrast cascade in $^{66}$Ge \cite{stefanova_66ge}. In addition, the standard deviation ($\sigma_{\Theta_{exp}}$~=~1.07) of the logarithmic $\beta$-decay-time distribution, which is obtained by gating on the 929-keV, 1135-keV and 1456-keV lines, meets the recommended limits ($\sigma_{\Theta_{exp}}^{\text{lower}}$~=~0.77, $\sigma_{\Theta_{exp}}^{\text{upper}}$~=~1.75) for 16 events~\cite{KHS2}. The derived $\beta$-decay half-life of 38$^{+13}_{-8}$~ms is also in agreement with Ref.~\cite{blank_66Se}. The similarity of the $\gamma$-ray energies with those of the isobaric partners $^{66}$As and $^{66}$Ge together with the arguments above, indicate that the observed $\gamma$ rays originate from $^{66}$Se.\par

\begin{figure}
\includegraphics[width=0.475\textwidth]{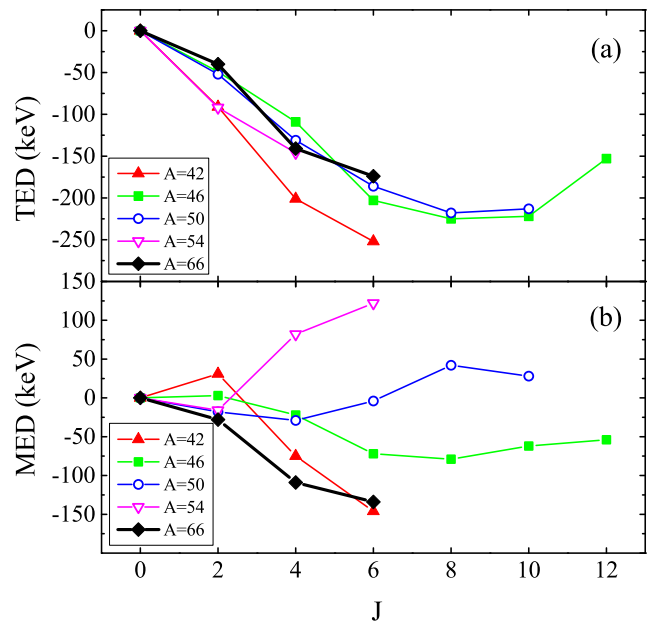}
\caption{\label{fig2} (Color online) (a) The TED for the nuclei between mass \textit{A}~=~42~$-$~66. (b) Same as (a) but for the MED. Data are taken from the present work and from Refs.~\cite{chiara_42Sc,endt_42Ti_42Ca,garrett_46V,garrett_46Cr,brandolini_46Ti,oleary_50Mn,lenzi_50Fe,brandolini_50Cr,rudolph_54Co,gadea_54Ni,rudolph_54Fe,rudi,stefanova_66ge}. Error bars are excluded for clarity.}
\end{figure}

The TED and MED data for \textit{A}~=~66 are plotted in Fig.~\ref{fig2}(a) and (b), respectively, along with the data for nuclei in the $f_{7/2}$ shell. The TED follow a negative trend within each triplet while the MED vary from case to case. The significant variation of the MED reflects the fact they depend strongly on Coulomb multipole effects associated with recoupling the angular momenta of pairs of particles as a function of spin. Thus the sign of the MED depends on whether it is protons or neutrons that are active in a particular member of the mirror pair. In addition, monopole effects will also contribute and again these will vary in sign from case to case. However, the TED are remarkably consistent in sign and, to a large extent, magnitude. This is partly associated with the fact that multipole effects will dominate the TED. Indeed, under the assumption of identical wave functions across the triplet, the monopole contributions discussed earlier would effectively cancel in the calculation of the TED. Identical wave functions is a reasonable assumption for well-bound states, although in heavier systems, there are predictions of different shape-driving effects that will destroy this symmetry~\cite{petrovici}.\par

\begin{figure}
\centering
\includegraphics[width=0.48\textwidth]{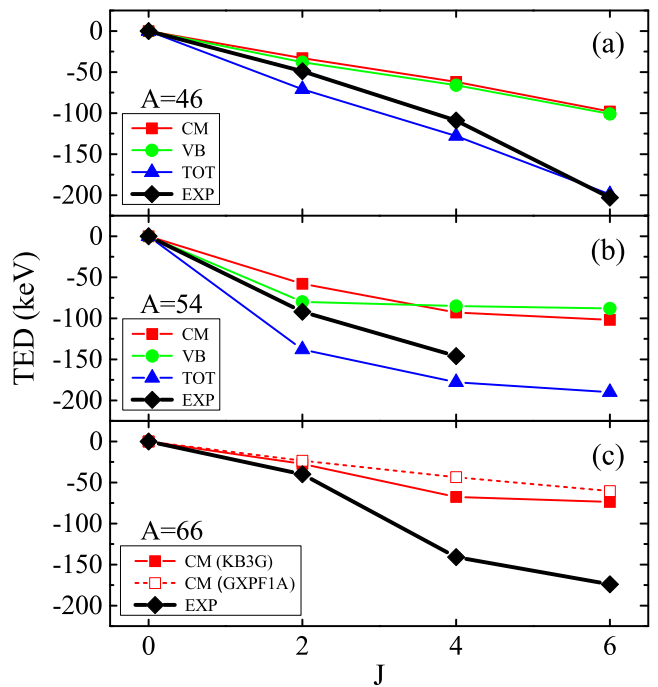}
\caption{\label{fig3} (Color online) The experimental and shell-model predicted TED for (a) \textit{A}~=~46, (b) \textit{A}~=~54 and (c) \textit{A}~=~66 triplets (see text for details).}
\end{figure}

The fact that the TED are negative can be explained in a simple picture when it is considered that the TED are directly dependent on the isotensor part of the two-body interaction - \textit{i.e.} $V_{\text{pp}}+V_{\text{nn}}-2V_{\text{np}}$ (compare with Eq.~\ref{TED}). The TED, thereby, depend on the difference between the np interaction and the average of the pp and nn interactions. The fact that the TED decrease with spin has its origin in two separate effects. Firstly, the number of \textit{T}~=~1 np pairs, for a given analogue state, is always larger in the odd-odd \textit{N}~=~\textit{Z} nucleus than in the two even-even nuclei. This has been demonstrated both analytically~\cite{engel} and with shell-model calculations in the $f_{7/2}$ shell~\cite{lenziPRC}. Secondly, the Coulomb isotensor interaction is positive, but reduces relative to the ground state for increasing angular momentum coupling. The combination of these two effects leads to the negative TED in all cases studied so far. However, in the $f_{7/2}$ shell, it was found that the Coulomb isotensor interaction (CM) alone was not sufficient to account for the TED magnitude~\cite{zuker,gadea_54Ni,bentley}. An additional nuclear isotensor component (VB) of +100~keV for \textit{J}~=~0 couplings of $f_{7/2}$ particles was identified based on the empirical TED of the \textit{A}~=~42 triplet~\cite{zuker}. The inclusion of this term gave a much better description of the TED in the shell-model prescription. This is also demonstrated in Fig.~\ref{fig3}(a) and (b) where the experimental and predicted TED are shown for states up to $6^+$ in the \textit{A}~=~46 and \textit{A}~=~54 triplets. The shell-model results presented have been performed using the procedure previously applied for these nuclei in Refs.~\cite{zuker,gadea_54Ni,bentley} using the code ANTOINE~\cite{antoine} and the KB3G interaction. The full \textit{fp} space was used for \textit{A}~=~46 and for \textit{A}~=~54 the number of excitations out of the $f_{7/2}$ shell is restricted to six. A total isotensor component of +100~keV is used, which is equivalent to making the np interaction 50~keV stronger than the average of pp and nn interactions. The results reproduced here are for completeness and comparison, and are the same as previously published for \textit{A}~=~46~\cite{zuker} and \textit{A}~=~54~\cite{bentley}. The inclusion of the VB term clearly leads to better agreement with the data in these cases. For \textit{A}~=~54, Gadea \textit{et al.,}~\cite{gadea_54Ni} showed that a reduced isotensor component of +50keV gave results that match the experimental data more closely.\par

It is obvious that in the case of the \textit{A}~=~66 triplet studied here, the consistent negative TED behaviour continues, as observed in the $f_{7/2}$ shell. In addition, it appears that the CM component alone will not account for the observed TED. This is illustrated in Fig.~\ref{fig3}(c), which shows a prediction of the TED for \textit{A}~=~66 assuming a Coulomb isotensor interaction alone. The calculation was performed using ANTOINE in the \textit{fp} space with KB3G and GXPF1A interactions, allowing at most five excitations beyond the $f_{7/2}$ and $p_{3/2}$ orbitals. This should be viewed as a simplistic calculation, as it does not include the $g_{9/2}$ orbit. The VB component has not been included since, unlike in the $f_{7/2}$ shell, we have no empirical estimate of the strength. Nevertheless, even this simple calculation shows that the Coulomb part alone is insufficient to explain the experimental TED magnitude. It should be noted that the missing $(g_{9/2})^2$ components in the wave functions would only change the prediction for the TED by virtue of the different spin-dependent changes of the Coulomb energy for $g_{9/2}$ wave functions compared with the \textit{fp} orbitals. It seems unlikely that this would be sufficient to account for the large TED seen at high spins.\par

In a recent theoretical study by Kaneko \textit{et al.,}~\cite{kanekoTDE} a shell-model analysis of displacement energies was performed in order to study the effect of INC nuclear forces. In particular, they considered triplet displacement energies for the ground states. In general they found that the agreement with the data was much improved in the $f_{7/2}$ shell, when the additional isotensor interaction of +165~keV for the \textit{J}~=~0, \textit{T}~=~1 coupling was introduced. It is interesting to note that the isotensor interaction used is larger than previously considered in this region. In addition, it was found in Ref.~\cite{kanekoTDE} that the INC forces were less important for nuclei in the upper \textit{fp} shell, which stands in contrast with the shell-model results presented in the current study.\par

In conclusion, excited states in the proton-rich nucleus $^{66}$Se have been identified using the recoil-$\beta$ tagging method. This data allows the TED across the \textit{A}~=~66 triplet to be examined for the first time. The observed TED mirrors that of the triplets in the $f_{7/2}$ shell. Shell-model calculations in the present work suggest that, in common with the $f_{7/2}$ shell, the Coulomb isotensor component alone is insufficient to account for the experimental TED, pointing to a continued need for an additional nuclear isospin non-conserving interaction. This conclusion is at variance with the recent theoretical work of Ref.~\cite{kanekoTDE} for nuclei above mass \textit{A}~=~60, and clearly necessitates that further experimental and theoretical studies, which include the $g_{9/2}$ orbit, are undertaken.\par

This work was supported by the Academy of Finland under the Finnish CoE Programme, by the EU-FP7-IA project ENSAR under grant number 262010 and by the UK STFC under grant number ST/J000051. The authors acknowledge the GAMMAPOOL European Spectroscopy Resource for the loan of germanium detectors. TG acknowledges the Academy of Finland (grant 131665). PR acknowledges the Magnus Ehrnrooth foundation for the support for this work.\par

\bibliographystyle{apsrev}
\bibliography{references_Se66}

\end{document}